\documentstyle[preprint,prl,aps]{revtex}
\begin{document}
\author{E. del Barco$^1$ \thanks{%
Permanent address: Dept. de F\'\i sica Fonamental, Universitat de Barcelona,
Diagonal 647, 08028 Barcelona, Spain}, N. Vernier$^1$, J.M. Hernandez$^2$,
J.Tejada$^2$, E.M. Chudnovsky$^3$, E. Molins$^4$ and G. Bellessa$^1$}
\address{$^1$ Laboratoire de Physique des Solides, B\^atiment 510,\\
Universit\'e Paris-Sud, 91405 Orsay, France}
\address{$^2$ Dept. de F\'{\i}sica Fonamental, Universitat de Barcelona, Diagonal\\
647, 08028\\
Barcelona, Spain}
\address{$^3$ Physics Department, CUNY Lehman College,\\
Bronx, NY 10468-1589, U.S.A.}
\address{$^4$ ICMAB (CSIC), Campus Univ.\\
Autonoma de Barcelona, 08193 Cerdanyola, Spain}
\title{Quantum Coherence in Fe$_8$ Molecular Nanomagnets}
\maketitle

\begin{abstract}
We report observation of coherent quantum oscilations in spin-10 Fe$_8 $
molecular clusters. The powder of magnetically oriented Fe$_8$ crystallites
was placed inside a resonator, in a dc magnetic field perpendicular to the
magnetization axis. The field dependence of the ac-susceptibility
was measured up to 5 T, at 680 MHz, down to 25 mK. Two peaks in the
imaginary part of the susceptibility have been detected, whose positions
coincide, without any fitting parameters, with the predicted two peaks
corresponding to the quantum splitting of the ground state in the magnetic field
parallel and perpendicular to the hard magnetization axis.
\end{abstract}

\newpage
The high value of the spin, $S=10$, in molecular clusters like Mn$_{12}$ and
Fe$_8$ allows to test the border between quantum and classical mechanics
\cite{Chudnovsky,Stamp,Schwarzchild}. For a large spin, the transitions
between degenerate spin levels appear in high-order of the perturbation theory
on the spin-phonon, crystal field, dipole, hyperfine, and other interactions.
This results in a long lifetime of spin states and produces a fascinating
opportunity to study the quantized spin levels and transitions between them
in macroscopic magnetization measurements
\cite{Paulsen,Novak,Friedman,Hernandez,Thomas,Luis,Sangregorio}.
Owing to a large anisotropy, the classical magnetic energy of the clusters
in zero magnetic field has two symmetric minima separated by the energy barrier.
From the classical point of view, the magnetic relaxation in Mn$_{12}$ and
Fe$_{8}$ crystals can be viewed as thermal activation over this barrier.
However, when the temperature becomes small compared to the anisotropy energy,
the relaxation is dominated by quantum tunneling under the barrier. The
corresponding matrix elements arise from the terms in the Hamiltonian which
do not commute with the equilibrium orientation of the spin
\cite{Korenblit,Schilling,Hemmen,CG,Garanin}.
If these terms are small, tunneling from the
levels near the bottom of the barrier is negligible but tunneling at
the top of the barrier may have a significant rate \cite{Garanin}.
It has been firmly established experimentally for both Mn${_{12}}$
\cite{book} and Fe$_8$ \cite{Sangregorio} that within a certain temperature
range the tunneling from thermally populated spin levels is
responsible for the magnetic relaxation. In Fe$_{8}$ the distance
between the ground state level and the first excited spin level is about 5K.
As the temperature is lowered well below that value, the populations of
excited spin levels subside to exponentially small values and only transitions
between the ground state levels become of practical interest \cite{Sangregorio}.
By applying the external dc magnetic field one can adjust the rate of these
transitions to the frequency of the ac field and observe coherent quantum
oscillations of the spin, similar to the textbook example of ammonia molecule.
For a macroscopic spin it has been attempted in antiferromagnetic particles
of ferritin \cite{Awschalom}, while for a small spin the coherent oscillations
have been reported in rare-earth ions glasses \cite{Vernier1} and non-oriented
CrNi$_{6}$ clusters \cite{Vernier2}

Among systems which consist of
well-characterized \cite{Barra,Sangregorio,Caciuffo} identical nanomagnets
Fe$_{8}$ has the highest spin and, therefore, the
ac-susceptibility study of the quantum splitting of its ground state can
be an important landmark in the search for macroscopic quantum coherence.
Besides its value for fundamental quantum physics, the observation of
quantum coherence in molecular nanomagnets may also add them to the list
of candidates for elements of quantum computers.
In this letter, we report such an experiment at 680 MHz down to 25 mK on
magnetically oriented Fe$_8$ grains in a static magnetic field perpendicular
to the easy magnetization axes of the grains. The advantage of this configuration
is the control of the tunneling rate by the magnetic field \cite{book}. We have
chosen Fe$_{8}$ over Mn$_{12}$ because in Mn$_{12}$ strong hyperfine fields
split each molecular spin state into a few hundred levels closely packed
into an energy band which is wider than the tunneling splitting
$\Delta$ \cite{Hartmann}. On the contrary, in Fe$_{8}$ the hyperfine fields
are very weak in the majority of clusters and the resulting splitting of the
molecular spin states remains within the limit $H_{z} < \Delta /2g \mu_{B}$ ($g$
being the gyromagnetic factor).

Pure Fe$_8$ crystals of length ranging from less than 1${\mu }$m to 2mm were
synthesized according to Ref. \cite{Wiedghardt}. The nominal composition, $%
(((C_6H_{15}N_3)_6Fe_8({\mu }_3-O)_2({\mu }_2-OH)_{12}(Br_7(H_2O))Br_8H_2O)$
, was checked by chemical and infrared analysis. The matrix orientation of the
crystals was performed by indexing 25 randomly searched reflections inside
the Enraf-Nonius CAD4 X-ray diffractometer with graphite monochromated MoK${%
\alpha }$ radiation. The measured crystall cell parameters, $a=10.609(7)$, $%
b=14.15(2)$, $c=15.002(9)$\AA , ${\alpha }=89.45(9)$, ${\beta }=10.03(5)$, ${%
\gamma }=109.42(9)\deg $, are in accordance with published values \cite
{Wiedghardt}. The Mossbauer spectrum of the crystals, in accordance with the
published data \cite{Barra}, evolves from an asymmetric paramagnetic doublet
at room temperature to three magnetic hyperfine sextets below 4 K. This
corresponds to the blocking of the trivalent Fe cations in three different
crystallographic sites inside the spin-10 Fe$_8$ cluster.

Before carrying out the high-frequency experiments, we also performed the dc
and ac magnetic characterization of our samples. Both, oriented single
crystals and oriented powder, have been studied. The orientation of single
crystals was done inside the Enraf-Nonius diffractometer, while the
orientation of the powder was done by solidifying an epoxy (Araldit) with Fe$%
_8$ micrometric crystallites buried inside, in a 5.5 T field at 290 K during
12 hours. The data on both, single crystals and oriented powder, are similar
to those obtained in Ref. \cite{Sangregorio}. Below the blocking
temperature, which depends logarithmically on the frequency of the ac-field,
periodic steps appear in both, in-phase and out-of-phase, components of the
susceptibility, with a period of 0.24 T. Fig. 1 shows the variation of the
out-of-phase component of the susceptibility versus magnetic field at 500 Hz and
2 K.

To measure the magnetic susceptibility at high frequency, we used a
split-ring resonator (also called loop-gap resonator) \cite{Hardy}. Its
frequency resonance was around 680 MHz and its quality factor was 3100 at
low temperature. Because the electric field exists in the gap and not in the
loop, this resonator is particularly attractive for magnetic susceptibility
measurements. The sample consisted of oriented Fe$_8$ micrometric crystallites
imbeded in an epoxy slab. The total mass of the crystallites
was 0.08 g. The external magnetic field was obtained from a superconducting
magnet. The resonator was pressed against the wall of the mixing chamber of
a $^3$He-$^4$He dilution refrigerator. The steady magnetic field was applied
perpendicular to the easy axis of the crystallites, whereas the ac field was
parallel to the latter. This was important for having non-vanishing matrix
elements between the two levels originating from the splitting of the
ground state. To measure the magnetic
susceptibility $\chi$, the resonance line of the resonator with the sample
inside was determined using electromagnetic pulses of low repetition rate
(to avoid heating of the sample). The line shape was Lorentzian. The line
broadering was proportional to the imaginary part of the susceptibility ($%
\chi$") and the resonance frequency shift was proportional to its real part,
($\chi$'). Fig.2 shows the variation of $\chi $'' as a function of the magnetic
field for two temperatures. Two peaks are clearly present, the first one at $%
H_1=2.25\pm 0.05$ T and the second one at $H_2=3.60\pm 0.05$ T. The peaks at
25 mK have roughly the same heights as the ones at 200 mK. At higher
temperature they broaden and disappear in the background noise.
The two peaks also exist, though are much less pronounced, for the field
parallel to the easy magnetization axis.

To the first approximation, the Hamiltonian of Fe$_8$ is \cite{Barra,Sangregorio}:
\begin{equation}
\label{hamiltonian}{\cal {H}}=-DS_z^2+ES_x^2-g\mu _B{\bf {H}\cdot {S}}
\end{equation}
Numerical diagonalization of this Hamiltonian with $D=0.31$ K and $E=0.092$
K (Ref. \cite{Barra}) shows that at a $2-3$ T field, directed perpendicular
to the z-axis, the splitting ${\Delta }$ of the two lowest states becomes of
the order of a few tens of mK, which limits $H_z$ by a few Gauss if one is
to look for the quantum coherence (the frequency of our resonator correponds
to an energy $\hbar \omega =$ 32 mK). The dipole fields can, in principle, be
greater. However, in the oriented zero-field-cooled sample the numbers of
spin looking "up" and "down" are equal. Statistically, about 1/8 of molecules
must be sensing zero dipole field.  The more ``dangerous'' $H_z$ field
is coming from the impossibility to apply the external field exactly perpendicular
to the easy axis. The above limitation on the longitudinal field would
require the accuracy in the orientation of the Fe$_8$ crystal with respect to the magnetic
field better than $H_z/H_x{\sim }10^{-4}$ rad, which is difficult to achieve.
For that reason, despite having at our disposal large single crystals of Fe$_8$,
we chose to work with the oriented powder of Fe$_8$ crystallites. If the
orientations of
the grains are within a cone of 0.1 rad, which is possible to achieve, the $%
10^{-6}$ fraction of the sample, that is to say a macroscopic number of Fe$%
_8 $ clusters, will satisfy the resonance condition. To see the
corresponding resonance in the ac-susceptibility measurements, the frequency
of the ac-field must equal ${\Delta }/{\hbar }$, which, for the above
numbers, corresponds to a few hundred MHz.

The above frequency must be greater than the frequency of the absorption and
emission of phonons, or other excitations, by the magnetic clusters,
otherwise the coherence will be destroyed. Although little is known about
the interaction of the clusters with the environment, it is believed that at
low temperature they are in the underdamped regime \cite{Leggett}. In that
regime the frequency in question is the pre-exponential factor (the attempt
frequency) of the tunneling rate. According to Ref. \cite{Sangregorio}, in Fe%
$_8$ this frequency is about 30 MHz.

When the field is applied perpendicular to the easy axis, the splitting ${%
\Delta }$ depends on the magnitude of the field $H$ and its angle ${\phi }$
with the hard axis. The dependence of ${\Delta }$ on $H$, for different
values of ${\phi }$, obtained by the numerical diagonalization of the
Hamiltonian (Eq. \ref{hamiltonian}) using $D=0.31$ K and $E=0.092$ K, is
plotted in Fig. 3. The insert shows $\Delta (\phi )$ for a fixed value of $H$%
. The pronounced minima at ${\phi }=0$ are due to the non-Kramers
topological quenching of tunneling noticed by Garg \cite{Garg}. In the
absence of dissipation, the contribution of each Fe$_8$ crystallite to the
imaginary part of the susceptibility ${\chi }"$ is proportional to ${\delta }%
({\omega }-{\Delta })$. At a given $H$, the total ${\chi }"$ is then
proportional to
\begin{equation}
\label{susceptibility}\int_0^\pi g({\phi }){\delta }({\omega }-{\Delta }[{%
\phi },H])d{\phi },
\end{equation}
where $g({\phi })$ is the distribution of crystallites over ${\phi }$. Since
no preferred orientation on the angle ${\phi }$ is expected, the integral in
Eq.\ref{susceptibility} is proportional to $|{\partial }{\Delta }/{\partial }%
{\phi }|^{-1}$. Now, we can notice that according to the inset of Fig. 3 the
derivative of ${\Delta }$ over ${\phi }$ equals zero at ${\phi }=0$ and ${%
\phi }={\pi }/2$. Therefore, we conclude that ${\chi }"(H)$ must have two
peaks, which are solutions of the equations:
\begin{equation}
\label{fields}
\begin{array}{c}
\Delta (
\frac \pi 2,H_1)=\omega  \\ \Delta (0,H_2)=\omega
\end{array}
\end{equation}
where ${\omega }=2{\pi }f$ and $f$ is the frequency of the ac field. That
is, the two resonance peaks we have experimentally observed correspond to
the quantum splitting of the ground state for the cases when the field is
perpendicular and parallel to the hard axis. From the positions of these two
peaks, we have extracted the values $D=0.275\pm 0.005$ K and $E=0.092\pm
0.005$ K of equation \ref{hamiltonian}. These values are in remarkable
agreement with the data given by other authors
\cite{Barra,Sangregorio,Caciuffo}. It remains to be explained, however,
why the same two peaks of much lower intensity appear in the field parallel to
the easy axis (Fig.2). In our opinion, this must be due to the non-perfect
orientation of the crystallites. If the orientation is done below 9T, there is
always a small fraction of the crystallites perpendicular to the field
independently of its direction \cite{Zhang}. It should also be mentioned that
we have not observed either spin echoes or the non-linear dependence of
the susceptibility on the power of the ac field \cite{Vernier1}. This can be
due to very short relaxations times. Indeed, the large  non-resonant
magnetic susceptibility suggests a large number of non-resonant
magnetic moments which should give spectral diffusion and, thus, reduce
efficiently the relaxation time of the resonant Fe$_8$ clusters.

The classical two states for which we observe the coherence are two symmetric
${\bf S}$ states at some angle with the applied dc field, shown in the insert
to Fig.4. Numerically, we find that the classical barrier height vanishes at
$H_{min}$ that depends on the angle betwen the field and the hard axis. This
dependence has two extrema: $H_{min}=3.34T$ and $H_{min}=4.71T$, which
roughly have the same ratio as the two experimental peak values. However,
the experimental peaks are located at 2.3 T and 3.6 T, where the barrier
still has the height of about 3 K, much higher than the experimental
temperature. It is also instructing to show quantum states which exhibit
quantum coherence. The gap $\Delta$  separates the first excited state $|1>$
from the ground state $|0>$. Each of these states can be written as a
superposition of the eigenstates $|m>$ of $S_{z}$:
\begin{equation}
|0>={\sum}_{m=-10}^{m=10}A_{m}|m>\,, |1>={\sum}_{m=-10}^{m=10}B_{m}|m>\,.
\end{equation}
Here $A_{m}=A_{-m}$ while $B_{m}=-B_{-m}$. Fig.4 shows $|A_m|^{2}$ and
$|B_m|^{2}$ for the first resonance, $H=2.25T$, as functions of $m$ for
$-10<m<10$. These are quantum counterparts of classical canted spin states
shown in the insert.

To conclude, we have observed coherent quantum oscillations in spin-10 Fe$_8$
molecular clusters. The two observed susceptibilty peaks are expected from
the biaxial system of particles oriented along the easy axis but radomly
distributed with respect to the orientation of their hard axes. The
quantum coherence explanation provides a remarkable agreement between the
parameters $D$ and $E$ of the Hamiltonian deduced from our experiment and
those previously reported.

J. T., and J. M. H. acknowledge support from the CICYT project No. IN96-0027
and the CIRIT project No. 1996-PIRB-00050. E. del B. acknowledges support
from the University of Barcelona. The work of E.M.C. has been supported by
the U.S. National Science Foundation Grant No. DMR-9024250.
The Laboratoire de Physique des Solides is associated with the C.N.R.S.

\pagebreak

\pagebreak

FIGURE CAPTIONS\\

{\bf Fig. 1}. Periodic steps in the out of phase low-frequency
ac-susceptibility of Fe$_8$ due to resonant spin tunneling, at 500 Hz and 2
K.

{\bf Fig. 2}. Dependence of the imaginary part of the high-frequency
ac-susceptibility of Fe$_8$ as a function of transverse magnetic field. The
dashed lines indicate the peak locations $H_1$ and $H_2$ obtained from the
Hamiltonian using the parameters $D=0.31$ K and $E=0.092$ K. The solid
circles are obtained with the magnetic field parallel to the easy axis. For
clarity, the curves are arbitrary translated vertically.

{\bf Fig. 3}. Dependence of the ground-state splitting as a function of
transverse field in Fe$_8$ for different orientations of this one in the
plane perpendicular to the easy axis. The insert shows the angular
dependence of the splitting at a fixed field.

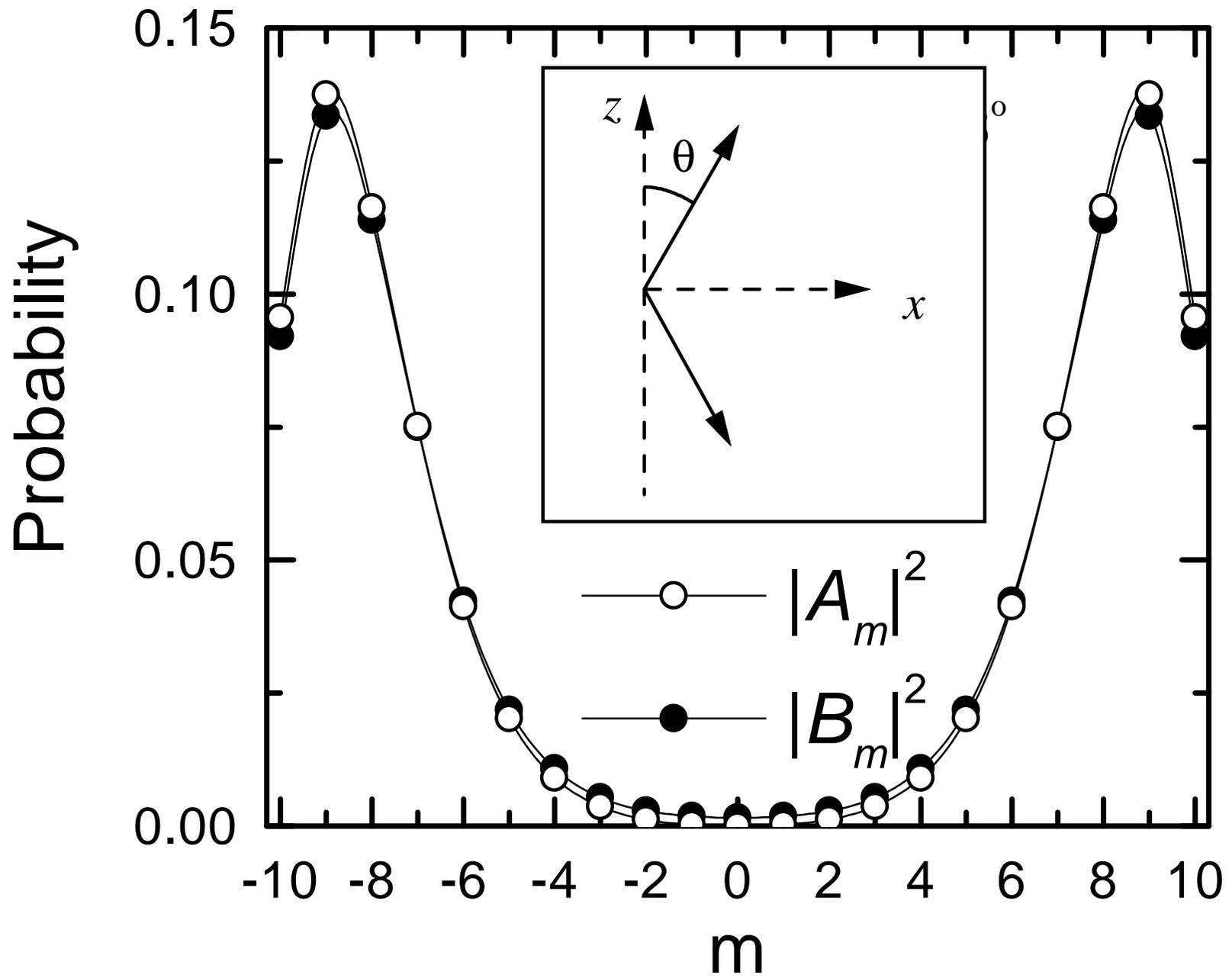
{\bf Fig. 4}. Probability distribution over $m$ in the ground state and the
first excited state, $|A_m|^{2}$ and $|B_m|^{2}$, respectively. The inset shows a
double-degenerate classical ground state.

\end{document}